\documentclass[preprint2]{aastex}

\usepackage{graphicx}
\usepackage{natbib}
\usepackage{float}
\usepackage{hyperref}

\shorttitle{Near real-time astrometry with the VLBA}
\shortauthors{Max-Moerbeck et al.}

\begin{document}

\title{Near real-time astrometry for spacecraft navigation with the VLBA:\\
A demonstration with the Mars Reconnaissance Orbiter and Odyssey}

\author{W. Max-Moerbeck,
W. F. Brisken,
and J. D. Romney}

\affil{National Radio Astronomy Observatory, P.O. Box 0, Socorro, NM 87801, USA}

\begin{abstract}
\noindent
We present a demonstration of near real-time spacecraft astrometry with the VLBA. We detect the X-band downlink signal from Mars Reconnaissance Orbiter and Odyssey with the VLBA and transmit the data over the internet for correlation at the VLBA correlator in near real-time.  Quasars near Mars in the plane of the sky are used as position references.  In the demonstration we were able to obtain initial position measurements within about 15 minutes of the start of the observation. The measured positions differ from the projected ephemerides by a few milliarcseconds, and the repeatability of the measurement is better than 0.3 milliarcseconds as determined from measurements from multiple scans. We demonstrate that robust and repeatable offsets are obtained even when removing half of the antennas. These observations demonstrate the feasibility of astrometry with the VLBA with a low latency and sub-milliarcsecond repeatability. 
\end{abstract}


\section{Introduction}

The Very Long Baseline Array (VLBA) is an interferometer of ten 25 meter diameter antennas observing from 300 MHz to 90 GHz. Eight of the ten antennas are distributed throughout the continental United States, one is on Mauna Kea, Hawaii and one is on St.\ Croix, Virgin Islands \citep[][]{napier1995}.  Long baselines (up to 8611 km) make it possible to obtain an image angular resolution of 0.1 by 0.2 milliarcseconds at 8.4~GHz, and typical relative astrometric accuracies of about 100 to 200 microarcseconds for standard phase-referencing observations \citep[e.g.,][]{loinard+2007, melis+2014}. 
These capabilities allow the study of non-thermal astronomical phenomena at small angular scales and provide precise astrometry, key to several applications \citep[][]{romney+reid2005, zensus+1998}. 
Precise astrometry has applications in spacecraft navigation as has been demonstrated by previous efforts by the National Radio Astronomy Observatory (NRAO) with the Spacecraft Navigation Pilot Project \citep[][]{romney2003,lanyi+2005,martin-mur+2006} and the Phoenix Mars lander experiment \citep[][]{fomalont+2010}. These techniques have also been used to measure the mass of Iapetus using data from a Cassini flyby \citep[][]{jacobson+2006}, observe Huygens as it entered Titan's atmosphere to measure its position and help determine wind information \citep[][]{bird+2005, folkner+2006, witasse+2006}, and to improve the ephemerides of Saturn by using Cassini observations \citep[][]{jones+2011}. All these projects used the standard VLBA observing mode in which data are recorded at the stations and shipped for correlation at the operations center in Socorro, NM. This process introduces a delay of about 2 weeks between observation and correlation of the data. 

The VLBA recently completed a sensitivity upgrade that included a new digital backend, a new Mark5C data recorder system and a new software correlator. The upgraded instrumentation at each VLBA antenna includes a computer in the digital data path which is primarily used for data routing to the recorder. This  provides a convenient point to access the data allowing a reduction in the latency of the data transfer from weeks to hours and thus enabling near real-time correlation. The correlation process has been simplified and made more flexible by the Distributed FX correlator \citep[DiFX,][]{deller+2007, deller+2012} which became the VLBA correlator in 2009 and has continued to evolve in capability since then. These recent upgrades have opened the possibility of doing near real-time observations with the VLBA, which is a very attractive possibility for time critical applications such as spacecraft navigation. To demonstrate the capabilities of the DiFX correlator for spacecraft navigation and exercise the use of the predicted ephemerides we performed observations of Voyager around the time it was believed to  enter interstellar space \citep[][]{burlaga+2013, krimigis+2013, stone+2013}.

Here we present the results of a recent effort to demonstrate the feasibility of near real-time astrometry of two spacecraft, Mars Reconnaissance Orbiter (MRO) and Odyssey, though passive observations of their X-band downlink signals. We give a description of the observations including the preparatory work required to identify suitable phase calibrators for relative astrometry. We also describe the setup we use for near real-time data transfer over the internet, and the correlation process with DiFX using predicted ephemerides for each spacecraft. The data reduction process and results of these observations are described, including the latency, offsets with respect to predicted orbits, their accuracy and repeatability, and the redundancy and tolerance to missing antennas for the astrometry. We close the paper with a discussion of some future developments required to make these observations standard and to improve the absolute astrometry.

\section{Observations}

This near real-time astrometry demonstration consists of observations at four epochs in 2013 September for the MRO and Odyssey observations and a preparatory observation of calibrators in 2013 August. 

\subsection{Demonstration observations}

The preparatory observation, performed on 2013 August 31, was a survey of possible phase calibrators near the position of Mars at the time of the near real-time astrometry demonstration. These data were used to determine the best and closest VLBA calibrators\footnote{\url{http://www.vlba.nrao.edu/astro/calib/}} within a set of candidate sources less than 2 degrees from Mars at the time of the near real-time demonstrations. These calibrators should be bright enough to be detected in a short integration, and be nearly point like to minimize phase errors. Each one of the candidate sources was used to produce model images to use during the phase calibration of the demonstration observations. These observations were performed with standard latency with bandwidths of 1 MHz and 16 MHz. All the  observations were reduced using standard methods for calibration and imaging in NRAO's Astronomical and Image Processing System (AIPS)\footnote{\url{http://www.aips.nrao.edu}}. The 16 MHz bands were used to make the model images and measure integrated flux densities and angular sizes. From these results we selected the most compact and brightest sources to be used as the calibrators, whose parameters are given in Table \ref{calibrators-summary}. The 1 MHz images for the selected candidates were used to determine the minimum integration times required to obtain good phase solutions with the CALIB task on AIPS. In all cases we found good phase solutions even with 30~s of integration, which determines the minimum time required for phase calibrator integrations.


The demonstration observations were performed on 2013 September 20, 21, 24 and 25. At each epoch we observed for 2 hours tracking Mars with both spacecraft in the same VLBA beam. Data were recorded in two 1 MHz bands in right circular polarization centered at the downlink frequency for each spacecraft, 8439.444446 MHz for MRO and 8406.851853 MHz for Odyssey. To minimize data transmission, we use the lowest recording bandwidth available of 1 MHz. This produces only 8 Mbps of data per antenna. We used simple phase-referenced observations by switching between calibrator (90 s of integration) and Mars (30 s of integration), with occasional observations of an astrometric check source. These astrometric check sources have a known position and are used to assess the astrometric repeatability and accuracy. We also observe a fringe finder source at the beginning, middle and end of the observations. For each one of the epochs the distance between source and calibrator was about 1.5, 1.4, 1.8 and 1.2 degrees. A fixed Doppler correction was used which caused the central frequency to drift by about 100 kHz, mainly as a result of the spacecraft orbital motion around Mars. This small drift is well within the observed 1 MHz band.

A short video made by the NRAO outreach office shows what happens in one of these demonstrations, and provides complementary information and a visual context to what is explained above\footnote{\url{http://vimeo.com/80099626}}.

\paragraph{Data transfer}

Most VLBA sites have limited network connectivity which is shared with monitor and control traffic from Socorro, NM. Two sites, Pie Town and Mauna Kea, can transfer at about 300 Mbps\footnote{Mbps: megabit per second, Gbps: gigabit per second} using 1 Gbps links funded by the US Naval Observatory (USNO) for their daily UT1$-$UTC observations. This opens the possibility of wide-band measurements for a single baseline. All other sites can transmit about 1.1 Mbps on 1.4 Mbps links which are the limiting factor on the latency. Thus we can transmit about 3 GB of data per antenna in 6 hours, which corresponds to 50 minutes of data at 8 Mbps and 12 seconds of data at the maximum recording rate of 2 Gbps.

The near real-time data transfer is enabled by the XCube switch\footnote{The XCube switch hardware is a product of the XCube Corporation \url{http://www.x3-c.com/}}, a Linux-PC based system situated in the data path from the digital backends and the Mark5C recorder\footnote{Haystack Observatory, Mark 5 memo \#57 \url{http://www.haystack.mit.edu/tech/vlbi/mark5/memo.html}}. For data rates up to about 500 Mbps the data can be recorded onto a standard random-access filesystem on an internal hard drive on the computer.
This is in contrast to the standard data path employing the Mark5C recorder where reading cannot be performed in parallel with writing. Files recorded to the internal hard drive are transmitted over the internet using the Tsunami-UDP file transfer program\footnote{\url{http://tsunami-udp.sourceforge.net}}. The ability to continue transfers after interruption and its high performance over long-haul switched networks leads to this program being commonly used for network transfer of Very Long Baseline Interferometry data.

\paragraph{Correlation}

There are several differences in the correlation of these observations compared to the vast majority of scientific observations. To accommodate the spacecraft which are swiftly moving, the VLBA DiFX correlator is fed with the most recent ephemerides from the NASA's Navigation and Ancillary Information Facility (NAIF)\footnote{NAIF: \url{http://naif.jpl.nasa.gov}}. The final positions determined do not depend in detail on the ephemeris; however, to prevent decorrelation (and hence loss of sensitivity) the ephemerides must predict the position sufficiently accurately (to a few arcseconds) as well as the rate of motion (to a few milliarcseconds per day). Because correlation occurred before the USNO published final values for Earth orientation parameters (EOP), predicted values had to be used. Since these parameters are used to connect the interferometer to the celestial reference frame, errors in these parameters result in errors in relative astrometry. Prediction errors for the EOP for a 5 day prediction horizon have a maximum error of 5 mas \citep[][]{malkim2000}. This results in a maximum offset error of about 87$\mu$as for a 1 degree distance between source and phase calibrator. The rms error of 2 mas for the same prediction horizon produces an rms offset error of about 35$\mu$as. The VLBA DiFX correlator uses GSFC Calc version 9.1 as the VLBI delay model with additional corrections for wavefront curvature and the solar system gravitational field to support near-field objects. The approximations made in the delay model start to break down at $\sim 10^6$ km distances. Two separate correlation passes were performed, one for each of the two spacecraft yielding two separate sets of correlator output that are stored in FITS-IDI format \citep[][]{greisen+2011}.

\subsection{Data calibration}

\begin{figure*}[h!]
\begin{center}
\includegraphics[angle=0, width=8cm, trim=0 0 0 0]{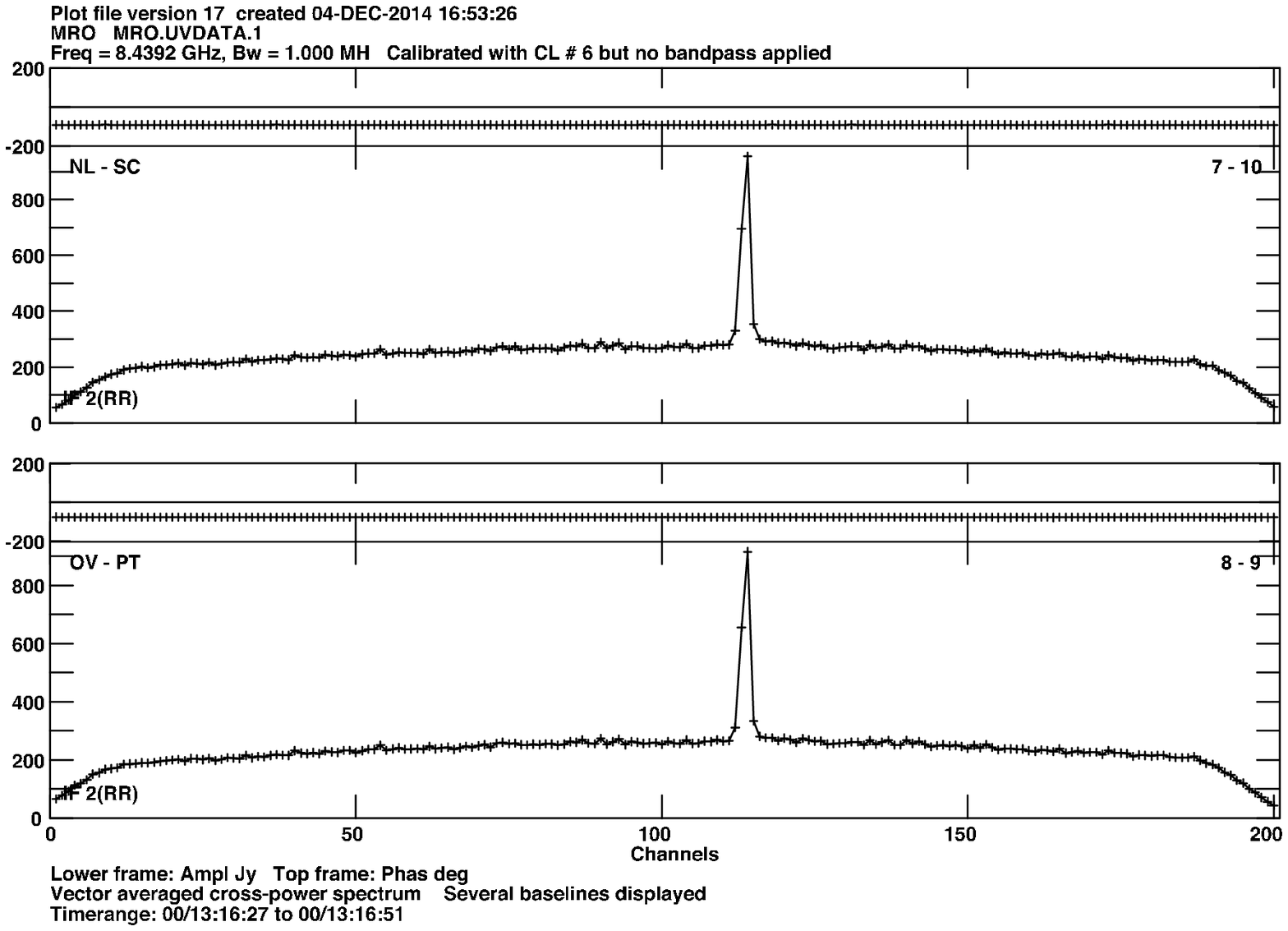}
\includegraphics[angle=0, width=8cm, trim=0 0 0 0]{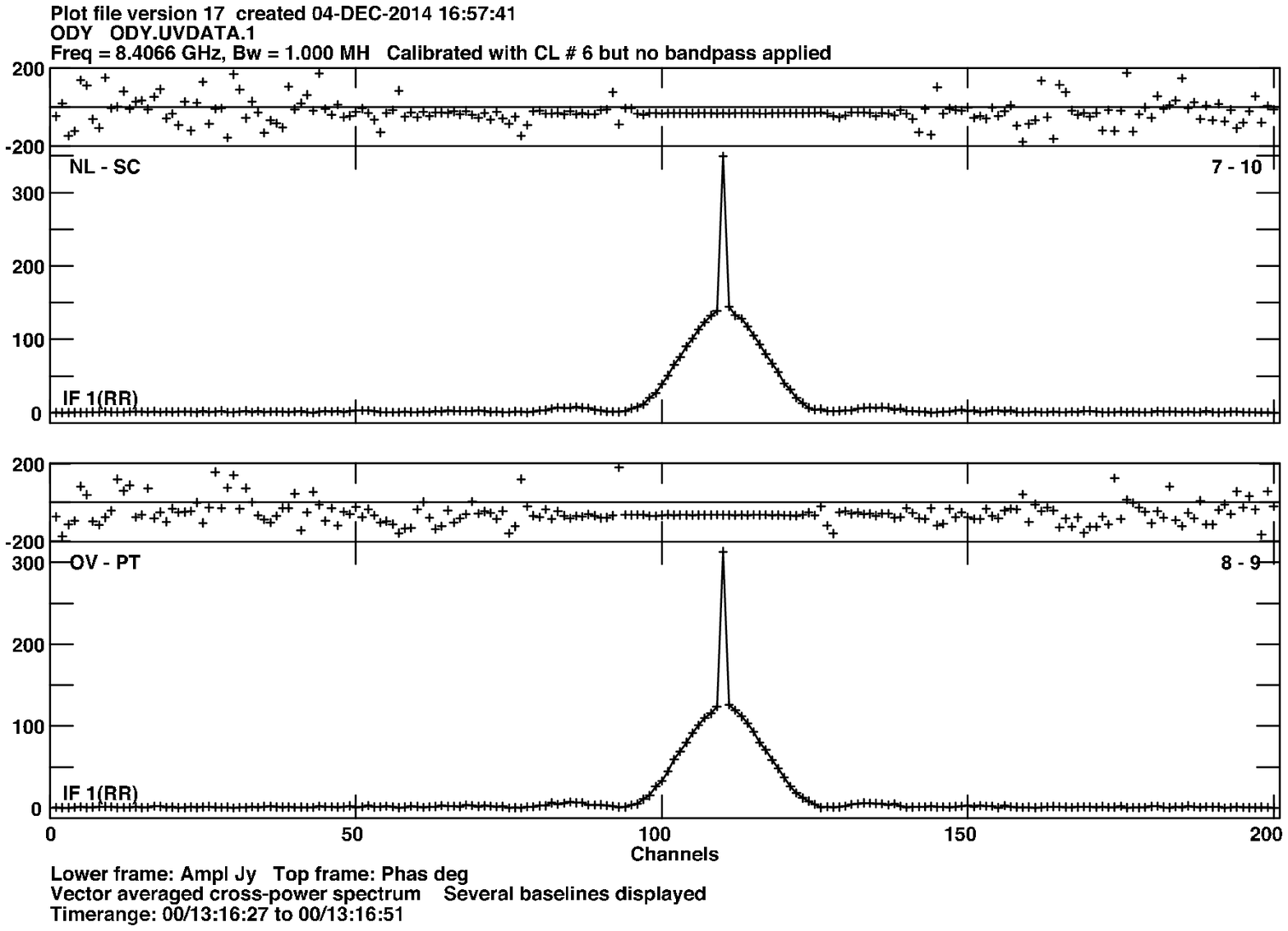}
\caption{Example of calibrated bandpass for MRO (right panels) and Odyssey (left panels). Each panel has phase in degrees on top and amplitude on the bottom, in kilo Jansky for MRO and Jansky for Odyssey, both as function of channel number. The upper panels are for the NL-SC baseline and the lower ones for the OV-PT baseline.}
\label{example-cal-bandpass}
\end{center}
\end{figure*}

Calibration using AIPS starts with amplitude calibration including digital sampler bias correction with the ACCOR task. During the near real-time observations no system temperature data are available until the end of the observing run, so system temperature calibration is skipped for the first stage of the data analysis. The data presented here include the system temperature calibration. This does not affect astrometry as it simply changes the absolute scale of the visibilities. Polarization parallactic angle rotation corrections are performed using the CLCOR task. Instrumental delay corrections are done by fringe fitting on a bright calibrator at the beginning of the observation using the FRING task. Instead of a bandpass calibration we simply exclude the 10\% outer channels, out of a total of 200, that have the largest gain variation. Phase calibration uses the nearby calibrator sources we modeled in the preparatory observations. For this last step we use the CALIB task. The phase calibration step registers the position of the spacecraft to the ICRF. Examples of calibrated bandpasses for the MRO and Odyssey are shown in Figure \ref{example-cal-bandpass}.


\subsection{Imaging and astrometry}

\begin{figure*}[ht!]
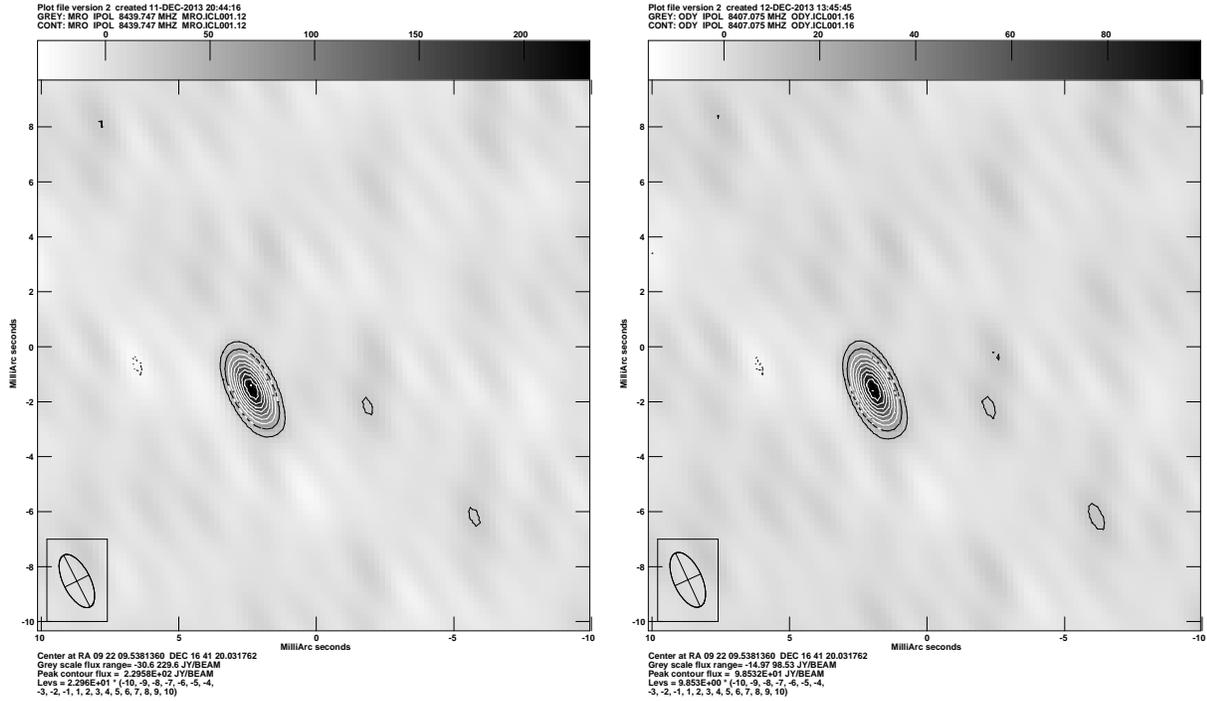

\begin{center}
\includegraphics[angle=0, width=8cm, trim=0 0 0 0]{fig2a.ps}
\includegraphics[angle=0, width=8cm, trim=0 0 0 0]{fig2b.ps}
\caption{Example of calibrated image for MRO (left panel) and Odyssey (right panel), for a simultaneous observation of both spacecraft on a single TS030E scan. Coordinates represent the offset with respect to the phase center used for correlation. The MRO offsets are $\Delta \alpha=2.34$ mas in right ascension and $\Delta \delta=-1.54$ mas in declination, with a signal to noise of 33.4. For the Odyssey the offsets are $\Delta  \alpha=1.93$ mas and $\Delta \delta=-1.57$ mas, with a signal to noise of 32.8. The artifacts in both images are almost identical due to correlated calibration errors arising from the close proximity of the sources, which are separated by at most 5 arcseconds over the different epochs and scans.}
\label{example-single-scan-image}
\end{center}
\end{figure*}

Imaging is done using the IMAGR task in AIPS. An image is produced for each 30 s scan first using all the available antennas but later discarding any problematic ones to improve the image quality. The images are 1024$\times$1024 pixels, with 0.2 mas pixels. For MRO we use all the channels, but for Odyssey only 60 channels centered on the peak are used. These channels are not the same for all scans due to Doppler shift that we are not compensating. Example images for a single simultaneous observation of MRO and Odyssey on a single TS030E scan are shown in Figure \ref{example-single-scan-image}. In the figure it can be seen that the artifacts in both images are almost identical due to correlated calibration errors arising from the close proximity of the sources, which are separated by at most 5 arcseconds over the different epochs and scans. This implies that our ability to measure the relative spacecraft positions is much better than their absolute ones (see the discussion in Section \ref{results}).


Astrometry is performed directly in the images by fitting a Gaussian model component with the JMFIT task which provides the offsets with respect to the phase center used for correlation.

\section{Results}
\label{results}

The main goal of these observations was to demonstrate the feasibility of near real-time astrometry and not to provide optimal results. With this limitation in mind it is still useful to look at the characteristics of the results that this low-bandwidth observations calibrated with a simple scheme can provide.

\subsection{Repeatability and accuracy}

We estimate the repeatability of the astrometry by obtaining the offsets between the center of the fitted Gaussian component and the phase center for each 30 s scan on each spacecraft. The offsets are $\Delta \alpha$ in R.A. and $\Delta \delta$ in decl. Since each offset is measured using non-overlapping data they provide independent measurements of the spacecraft position. The error budget for each measured offset will contain contributions from the ephemeris, calibrator position errors and calibration transfer errors. A summary of the repeatability results for the MRO is shown in Table \ref{repeatability-summary-mro} and for the Odyssey in Table \ref{repeatability-summary-ody}.


From these results we can see that the measured spacecraft position was offset from the ephemeris by different amounts on each observation, with values up to a couple of milliarcseconds. The scatter in the measured offsets is less than about 0.3 mas and in many cases less that 0.1 mas.

We estimate the accuracy of the astrometry by using the astrometric check source observations, which are observed in the same way as the spacecraft, but for a source of well determined position. A summary of the offset between the catalog and observed position of the calibrators is shown in Table \ref{accuracy-summary}. This offset can be as high as about 3~mas. This result indicates that the limiting factor for the accuracy is the quasar grid. This is a problem that could be minimized by obtaining astrometric observations on the calibrator sources.


We also estimate the repeatability of the relative offset between the spacecraft by taking the difference between estimated offsets for simultaneous scans in Odyssey and MRO. As pointed out in Figure \ref{example-single-scan-image} and associated discussion, these relative offset should have a lower scatter than each individual spacecraft offset. The standard deviation in the offset difference is less than 0.055 mas in all cases, and is much better in most cases. See Table \ref{repeatability-summary-relative-ody-mro} for a summary.


\subsection{Effect of reduced number of antennas}

The ten VLBA antennas\footnote{We refer to them by their code names: BR, FD, HN, KP, LA, MK, NL, OV, PT and SC.} can provide a highly robust system due to redundancy, which is key for applications requiring near real-time astrometry. In order to quantify the robustness of our results, we tested the effects in the astrometry of the loss of one or several antennas. This is tested by excluding sets of antennas from a single TS030C scan and repeating the same astrometric measurements performed on the complete data for each scan. Only 9 antennas were used on the complete data set because KP had sensitivity problems at the time of observations. Omitting a single antenna (i.e., using 8 antennas) resulted in good quality images for the 9 cases we tested with rms offsets with respect to the original image of 35 $\mu$as in $\alpha$ and 70 $\mu$as in $\delta$. The maximum excursion was about 130 $\mu$as, measured when removing SC. The second largest excursion, about 100 $\mu$as, was obtained when removing BR. We tested by omitting selected pairs of antennas (i.e., using 7 antennas), and measured maximum excursions of about 200 $\mu$as. Even larger offsets (up to $600\,\mu$as) are observed when including only 5 antennas (BR, FD, OV, LA, PT). These differences can be reduced by the use of improved calibration techniques and by increasing the time on source.

One should notice that LA experienced technical difficulties (excluded from TS030D/E). For TS030D, MK experienced high system temperatures during the second half of the epoch and was excluded. In the same epoch SC and HN were observing at elevations lower than 20 degrees and were also excluded. This resulted in TS030D effectively using 5 antennas, and producing lower quality results than the other epochs.

\section{Summary and discussion}

We presented the results of a demonstration of near real-time astrometry for the MRO and the Odyssey performed at four epochs during 2013 September. The feasibility of this observing mode was successfully demonstrated, with first detection of the fringe calibrator in less than 5 minutes after the start of the observations and first spacecraft offsets within 15 minutes of the start of the observations. The results are promising, even with the simple observation, calibration, imaging and astrometric procedures used, indicating that significant improvements can be made by refining these procedures. The measured positions differ from the projected ephemerides by a few milliarcseconds, and the repeatability of the measurement within a single epoch is better than 0.3 milliarcseconds as determined from measurements from multiple scans.

The technique is promising for spacecraft above $-30$ degrees of declination, beyond which calibration and mutual visibility of the constituent antennas become increasingly difficult to handle. The limited network connectivity limits the choice of calibrators to bright sources of at least 50 mJy for the current 1 MHz bandwidth. The image plane analysis we performed requires a good correlator model. Total delays are obtained as part of the data analysis and are readably obtainable.

Overall the demonstration was successful, providing results with good accuracy, repeatability and tolerance to missing antennas. Straightforward improvements to this technique can make it a valuable addition for spacecraft navigation.

\acknowledgements
W.M. thanks Amy Mioduszewski for useful conversations about phase-referenced observations with the VLBA. The National Radio Astronomy Observatory is a facility of the National Science Foundation operated under cooperative agreement by Associated Universities, Inc. This work made use of the Swinburne University of Technology software correlator, developed as part of the Australian Major National Research Facilities Programme and operated under licence. We thank the referee for comments that improved the presentation of the paper.

{\it Facilities:} \facility{VLBA}

\bibliographystyle{apj}
\bibliography{ms}

\begin{thebibliography}{}
\expandafter\ifx\csname natexlab\endcsname\relax\def\natexlab#1{#1}\fi

\bibitem[{{Bird} {et~al.}(2005){Bird}, {Allison}, {Asmar}, {Atkinson},
  {Avruch}, {Dutta-Roy}, {Dzierma}, {Edenhofer}, {Folkner}, {Gurvits},
  {Johnston}, {Plettemeier}, {Pogrebenko}, {Preston}, \& {Tyler}}]{bird+2005}
{Bird}, M.~K., {Allison}, M., {Asmar}, S.~W., {et~al.} 2005, \nat, 438, 800

\bibitem[{Burlaga {et~al.}(2013)Burlaga, Ness, \& Stone}]{burlaga+2013}
Burlaga, L.~F., Ness, N.~F., \& Stone, E.~C. 2013, Science, 341, 147

\bibitem[{{Deller} {et~al.}(2007){Deller}, {Tingay}, {Bailes}, \&
  {West}}]{deller+2007}
{Deller}, A.~T., {Tingay}, S.~J., {Bailes}, M., \& {West}, C. 2007, \pasp, 119,
  318

\bibitem[{{Deller} {et~al.}(2011){Deller}, {Brisken}, {Phillips}, {Morgan},
  {Alef}, {Cappallo}, {Middelberg}, {Romney}, {Rottmann}, {Tingay}, \&
  {Wayth}}]{deller+2012}
{Deller}, A.~T., {Brisken}, W.~F., {Phillips}, C.~J., {et~al.} 2011, \pasp,
  123, 275

\bibitem[{{Folkner} {et~al.}(2006){Folkner}, {Asmar}, {Border}, {Franklin},
  {Finley}, {Gorelik}, {Johnston}, {Kerzhanovich}, {Lowe}, {Preston}, {Bird},
  {Dutta-Roy}, {Allison}, {Atkinson}, {Edenhofer}, {Plettemeier}, \&
  {Tyler}}]{folkner+2006}
{Folkner}, W.~M., {Asmar}, S.~W., {Border}, J.~S., {et~al.} 2006, Journal of
  Geophysical Research (Planets), 111, 7

\bibitem[{{Fomalont} {et~al.}(2010){Fomalont}, {Martin-Mur}, {Border},
  {Naudet}, {Lanyi}, {Romney}, {Dhawan}, \& {Geldzahler}}]{fomalont+2010}
{Fomalont}, E., {Martin-Mur}, T., {Border}, J., {et~al.} 2010, in 10th European
  VLBI Network Symposium and EVN Users Meeting: VLBI and the New Generation of
  Radio Arrays

\bibitem[{{Greisen}(2011)}]{greisen+2011}
{Greisen}, E.~W. 2011, AIPS Memo Series, 114

\bibitem[{{Jacobson} {et~al.}(2006){Jacobson}, {Antreasian}, {Bordi},
  {Criddle}, {Ionasescu}, {Jones}, {Mackenzie}, {Meek}, {Parcher}, {Pelletier},
  {Owen}, {Roth}, {Roundhill}, \& {Stauch}}]{jacobson+2006}
{Jacobson}, R.~A., {Antreasian}, P.~G., {Bordi}, J.~J., {et~al.} 2006, \aj,
  132, 2520

\bibitem[{{Jones} {et~al.}(2011){Jones}, {Fomalont}, {Dhawan}, {Romney},
  {Folkner}, {Lanyi}, {Border}, \& {Jacobson}}]{jones+2011}
{Jones}, D.~L., {Fomalont}, E., {Dhawan}, V., {et~al.} 2011, \aj, 141, 29

\bibitem[{{Krimigis} {et~al.}(2013){Krimigis}, {Decker}, {Roelof}, {Hill},
  {Armstrong}, {Gloeckler}, {Hamilton}, \& {Lanzerotti}}]{krimigis+2013}
{Krimigis}, S.~M., {Decker}, R.~B., {Roelof}, E.~C., {et~al.} 2013, Science,
  341, 144

\bibitem[{{Lanyi} {et~al.}(2005){Lanyi}, {Border}, {Benson}, {Dhawan},
  {Fomalont}, {Martin-Mur}, {McElrath}, {Romney}, \& {Walker}}]{lanyi+2005}
{Lanyi}, G., {Border}, J., {Benson}, J., {et~al.} 2005, Interplanetary Network
  Progress Report, 162, A1

\bibitem[{{Loinard} {et~al.}(2007){Loinard}, {Torres}, {Mioduszewski},
  {Rodr{\'{\i}}guez}, {Gonz{\'a}lez-L{\'o}pezlira}, {Lachaume}, {V{\'a}zquez},
  \& {Gonz{\'a}lez}}]{loinard+2007}
{Loinard}, L., {Torres}, R.~M., {Mioduszewski}, A.~J., {et~al.} 2007, \apj,
  671, 546

\bibitem[{{Ma} {et~al.}(2009){Ma}, {Arias}, {Bianco}, {Boboltz}, {Bolotin},
  {Charlot}, {Engelhardt}, {Fey}, {Gaume}, {Gontier}, {Heinkelmann}, {Jacobs},
  {Kurdubov}, {Lambert}, {Malkin}, {Nothnagel}, {Petrov}, {Skurikhina},
  {Sokolova}, {Souchay}, {Sovers}, {Tesmer}, {Titov}, {Wang}, {Zharov},
  {Barache}, {Boeckmann}, {Collioud}, {Gipson}, {Gordon}, {Lytvyn},
  {MacMillan}, \& {Ojha}}]{ma+2009}
{Ma}, C., {Arias}, E.~F., {Bianco}, G., {et~al.} 2009, IERS Technical Note, 35,
  1

\bibitem[{{Malkin}(2000)}]{malkim2000}
{Malkin}, Z. 2000, in Astronomical Society of the Pacific Conference Series,
  Vol. 208, IAU Colloq. 178: Polar Motion: Historical and Scientific Problems,
  ed. S.~{Dick}, D.~{McCarthy}, \& B.~{Luzum}, 505

\bibitem[{{Mart\'in-Mur} {et~al.}(2006){Mart\'in-Mur}, {Antreasian}, {Border},
  {Benson}, {Dhawan}, {Fomalont}, {Graat}, {Jacobson}, {Lanyi}, {McElrath},
  {Romney}, \& {Walker}}]{martin-mur+2006}
{Mart\'in-Mur}, T., {Antreasian}, P., {Border}, J., {et~al.} 2006, in
  Proceedings of the Twenty-fifth International Symposium on Space Technology
  and Science, 587--592

\bibitem[{{Melis} {et~al.}(2014){Melis}, {Reid}, {Mioduszewski}, {Stauffer}, \&
  {Bower}}]{melis+2014}
{Melis}, C., {Reid}, M.~J., {Mioduszewski}, A.~J., {Stauffer}, J.~R., \&
  {Bower}, G.~C. 2014, Science, 345, 1029

\bibitem[{{Napier}(1995)}]{napier1995}
{Napier}, P.~J. 1995, in Astronomical Society of the Pacific Conference Series,
  Vol.~82, Very Long Baseline Interferometry and the VLBA, ed. J.~A. {Zensus},
  P.~J. {Diamond}, \& P.~J. {Napier}, 59

\bibitem[{{Romney} \& {Reid}(2005)}]{romney+reid2005}
{Romney}, J., \& {Reid}, M., eds. 2005, Astronomical Society of the Pacific
  Conference Series, Vol. 340, {Future Directions in High Resolution Astronomy:
  The 10th Anniversary of the VLBA}

\bibitem[{{Romney}(2003)}]{romney2003}
{Romney}, J.~D. 2003, in Bulletin of the American Astronomical Society,
  Vol.~35, American Astronomical Society Meeting Abstracts, 1268

\bibitem[{{Stone} {et~al.}(2013){Stone}, {Cummings}, {McDonald}, {Heikkila},
  {Lal}, \& {Webber}}]{stone+2013}
{Stone}, E.~C., {Cummings}, A.~C., {McDonald}, F.~B., {et~al.} 2013, Science,
  341, 150

\bibitem[{{Witasse} {et~al.}(2006){Witasse}, {Lebreton}, {Bird}, {Dutta-Roy},
  {Folkner}, {Preston}, {Asmar}, {Gurvits}, {Pogrebenko}, {Avruch}, {Campbell},
  {Bignall}, {Garrett}, {van Langevelde}, {Parsley}, {Reynolds}, {Szomoru},
  {Reynolds}, {Phillips}, {Sault}, {Tzioumis}, {Ghigo}, {Langston}, {Brisken},
  {Romney}, {Mujunen}, {Ritakari}, {Tingay}, {Dodson}, {van't Klooster},
  {Blancquaert}, {Coustenis}, {Gendron}, {Sicardy}, {Hirtzig}, {Luz}, {Negrao},
  {Kostiuk}, {Livengood}, {Hartung}, {de Pater}, {{\'A}d{\'a}mkovics},
  {Lorenz}, {Roe}, {Schaller}, {Brown}, {Bouchez}, {Trujillo}, {Buratti},
  {Caillault}, {Magin}, {Bourdon}, \& {Laux}}]{witasse+2006}
{Witasse}, O., {Lebreton}, J.-P., {Bird}, M.~K., {et~al.} 2006, Journal of
  Geophysical Research (Planets), 111, 7

\bibitem[{{Zensus} {et~al.}(1998){Zensus}, {Taylor}, \& {Wrobel}}]{zensus+1998}
{Zensus}, J.~A., {Taylor}, G.~B., \& {Wrobel}, J.~M., eds. 1998, Astronomical
  Society of the Pacific Conference Series, Vol. 144, {Radio Emission from
  Galactic and Extragalactic Compact Sources, IAU Colloquium 164}

\end{thebibliography}


\newpage

\begin{deluxetable}{c c c c c c c c c}
\tabletypesize{\scriptsize}
\tablecaption{Properties of the calibrator sources.\label{calibrators-summary}}
\tablewidth{0pt}
\tablehead{
\colhead{Source} & \colhead{RA} & \colhead{Dec} &  \colhead{RA error} &  \colhead{Dec error} & \colhead{Size\tablenotemark{a}} & \colhead{Peak flux} & \colhead{Note\tablenotemark{b}} & \colhead{Reference\tablenotemark{c}} \\
\colhead{} & \colhead{hh:mm:ss} & \colhead{dd:mm:ss} & \colhead{mas} & \colhead{mas} & \colhead{mas$\times$mas} & \colhead{mJy/beam} & \colhead{} & \colhead{}
}
\startdata
J0927+3902 & 09:27:03.0139380 & 39:02:20.851770  & 0.130 & 0.100 & 0.74$\times$0.42 & 6896$\pm$19 & Fringe & GSFC \\ 
J0908+1609 & 09:08:55.9253500 & 16:09:54.763880  & 0.553 & 1.393 & 0.36$\times$0.00 & 195$\pm$1 & Phase/Check & GSFC \\
J0909+1928 &	09:09:37.4434660 & 19:28:08.291220 & 0.140 & 0.190 & 0.74$\times$0.00 & 195$\pm$1 & Check & PETR \\
J0925+1658 & 09:25:49.9644620 & 16:58:12.203380  & 1.472 & 2.367 & 0.08$\times$0.00 & 148$\pm$1 & Phase & GSFC
\enddata

\tablenotetext{a}{This is the deconvolved source size major and minor axes. A size of zero indicates the source is unresolved in this direction.}

\tablenotetext{b}{``Fringe'' is for fringe fitting calibrator. ``Phase'' is for phase calibrator. ``Check'' is for astrometric check source.}

\tablenotetext{c}{GSFC: \citet[][]{ma+2009}\\ PETR: L. Petrov, solution rfc\_2012b (unpublished, available on the Web at \url{http://astrogeo.org/vlbi/solutions/rfc_2012b})}
\end{deluxetable}

\begin{deluxetable}{c c c c c c c}
\tabletypesize{}
\tablecaption{Summary of the repeatability of the offset measurements for MRO.\label{repeatability-summary-mro}}
\tablewidth{0pt}
\tablehead{
\colhead{Obs.} & \colhead{Date} & \colhead{Scans} & \colhead{$\overline{\Delta \alpha}$} & \colhead{$\overline{\Delta \delta}$} & \colhead{$\sigma_{\Delta \alpha}$} & \colhead{$\sigma_{\Delta \delta}$} \\
\colhead{} & \colhead{} & \colhead{[count]} & \colhead{[mas]} & \colhead{[mas]} & \colhead{[mas]} & \colhead{[mas]}
}
\startdata
TS030B & Sep 20 & 8 & 0.82 & 0.12 & 0.12 & 0.15 \\
TS030C & Sep 21 & 12 & 0.76 & 0.06 & 0.06 & 0.13 \\
TS030D & Sep 24 & 9 & 2.95 & $-$1.31 & 0.23 & 0.18 \\
TS030E & Sep 25 & 17 & 2.42 & $-$1.47 & 0.14 & 0.06
\enddata
\end{deluxetable}

\begin{deluxetable}{c c c c c c c}
\tabletypesize{}
\tablecaption{Summary of the repeatability of the offset measurements for Odyssey.\label{repeatability-summary-ody}}
\tablewidth{0pt}
\tablehead{
\colhead{Obs.} & \colhead{Date} & \colhead{Scans} & \colhead{$\overline{\Delta \alpha}$} & \colhead{$\overline{\Delta \delta}$} & \colhead{$\sigma_{\Delta \alpha}$} & \colhead{$\sigma_{\Delta \delta}$} \\
\colhead{} & \colhead{} & \colhead{[count]} & \colhead{[mas]} & \colhead{[mas]} & \colhead{[mas]} & \colhead{[mas]}
}
\startdata
TS030B & Sep 20 & 4 & 0.38 & 0.33 & 0.10 & 0.07 \\
TS030C & Sep 21 & 6 & 0.26 & 0.23 & 0.19 & 0.27 \\
TS030D & Sep 24 & 10 & 2.22 & $-$1.08 & 0.23 & 0.15 \\
TS030E & Sep 25 & 12 & 1.88 & $-$1.39 & 0.09 & 0.13
\enddata
\end{deluxetable}

\begin{deluxetable}{l c c c c c }
\tabletypesize{}
\tablecaption{Summary of the offset between catalog and observed calibrator position.\label{accuracy-summary}}
\tablewidth{0pt}
\tablehead{
\colhead{Obs.} & \colhead{Calibrator} & \colhead{Check source} & \colhead{Separation} & \colhead{$\Delta \alpha$} & \colhead{$\Delta \delta$} \\
\colhead{} & \colhead{} & \colhead{} & \colhead{[deg]} & \colhead{[mas]} & \colhead{[mas]}
}
\startdata
TS030B & J0908+1609 & J0909+1928 & 3.3 & 0.19 & $-$1.14 \\
TS030C & J0908+1609 & J0909+1928 & 3.3 & 0.23 & $-$1.24 \\
TS030D & J0925+1658 & J0908+1609 & 4.1 & 2.93 & $-$1.54 \\
TS030E & J0925+1658 & J0908+1609 & 4.1 & 1.78 & $-$1.78
\enddata
\end{deluxetable}

\begin{deluxetable}{c c c c c c c}
\tabletypesize{}
\tablecaption{Summary of the repeatability of the offset measurements difference for Odyssey and MRO.\label{repeatability-summary-relative-ody-mro}}
\tablewidth{0pt}
\tablehead{
\colhead{Obs.} & \colhead{Date} & \colhead{Scans} & \colhead{$\overline{\Delta \alpha}$} & \colhead{$\overline{\Delta \delta}$} & \colhead{$\sigma_{\Delta \alpha}$} & \colhead{$\sigma_{\Delta \delta}$} \\
\colhead{} & \colhead{} & \colhead{[count]} & \colhead{[mas]} & \colhead{[mas]} & \colhead{[mas]} & \colhead{[mas]}
}
\startdata
TS030B & Sep 20 & 2 & $-$0.170 & 0.163 & 0.027 & 0.016 \\
TS030C & Sep 21 & 4 & $-$0.651 & 0.243 & 0.054 & 0.045 \\
TS030D & Sep 24 & 8 & $-$0.711 & 0.219 & 0.025 & 0.009 \\
TS030E & Sep 25 & 9 & $-$0.405 & 0.020 & 0.007 & 0.055
\enddata
\end{deluxetable}

\clearpage


\newpage

\end{document}